# Controllable dimensionality conversion between 1D and 2D CrCl$_3$ magnetic nanostructures

Shuangzan Lu[1,4#], Deping Guo[2,5#], Zhengbo Cheng[1#], Yanping Guo[1#], Cong Wang[2,5], Jinghao Deng[1], Yusong Bai[1], Cheng Tian[6], Linwei Zhou[2,5], Youguo Shi[6], Jun He[1, 3*], Wei Ji[2,5*], Chendong Zhang[1*]

[1]School of Physics and Technology, Wuhan University, Wuhan 430072, China
[2]Department of Physics and Beijing Key Laboratory of Optoelectronic Functional Materials and Micro-Nano Devices, Renmin University of China, Beijing 100872, China
[3]Wuhan Institute of Quantum Technology, Wuhan 430206, China
[4]Hubei Jiufengshan Laboratory, Wuhan 430074, China
[5]Key Laboratory of Quantum State Construction and Manipulation (Ministry of Education), Renmin University of China, Beijing, 100872, China
[6]Beijing National Laboratory for Condensed Matter Physics and Institute of Physics, Chinese Academy of Sciences, Beijing 100190, China

[#]These authors contribute equally to this work.
*Correspondence and requests for materials should be addressed to:
He-jun@whu.edu.cn (J. H.), wji@ruc.edu.cn (W. J.), cdzhang@whu.edu.cn (C.D. Z)

**Abstract:** The fabrication of one-dimensional (1D) magnetic systems on solid surfaces, although of high fundamental interest, has yet to be achieved for a crossover between two-dimensional (2D) magnetic layers and their associated 1D spin chain systems. In this study, we report the fabrication of 1D single-unit-cell-width CrCl$_3$ atomic wires and their stacked few-wire arrays on the surface of a van der Waals (vdW) superconductor NbSe$_2$. Scanning tunneling microscopy/spectroscopy and first-principles calculations jointly revealed that the single wire shows an antiferromagnetic large-bandgap semiconducting state in an unexplored structure different from the well-known 2D CrCl$_3$ phase. Competition among the total energies and nanostructure-substrate interfacial interactions of these two phases result in the appearance of the 1D phase. This phase was transformable to the 2D phase either prior to or after the growth for *in situ* or *ex situ* manipulations, in which the electronic interactions at the vdW interface play a nontrivial role that could regulate the dimensionality conversion and structural transformation between the 1D-2D CrCl$_3$ phases.

Tailoring two-dimensional (2D) materials into one-dimensional (1D) structures such as nanoribbons and nanowires has been a prevailing approach for the manufacture of emergent properties in van der Waals (vdW) structures. Extensive attempts have been made to achieve controlled fabrication of the 1D counterparts of graphene [1-3] and semiconducting transition metal chalcogenides [4-6], which have exhibited rich electron correlated behaviors [5, 6]. The recent discovery of long-range magnetism in atomically thin vdW materials, *e.g.*, chromium trihalides, is a major breakthrough in the field of 2D materials. This discovery has opened up a new avenue for the applications of vdW structures in spintronics [7-14]. The integration of 1D and 2D magnetic nanostructures appears to be a promising strategy for minimizing the size of spintronic devices [15, 16]. The direct growth of 1D nanostructures on 2D layers is a straightforward strategy for 1D-2D vdW integration, which naturally raises the question of whether the 1D counterparts of these emerging 2D magnetic materials can be obtained on 2D substrates.

One-dimensional magnetic systems, usually in the form of spin chains, are themselves a thriving field of research and have been used to illustrate many key concepts in many-body quantum physics [17] and even triggered the recent search for new phases of matter (such as the Majorana fermion) [18-21]. Note that in a practical prototypic system, spin-chain excitations strongly depend on the coupling with the environment, mostly the holding substrate. Various ways exist to create physical realizations of spin chains on solid surfaces, based on either compound solids, molecular self-assembly, or trapped atoms [21-25]. However, a spin chain with vdW nature, which offers considerably magnetic exchange interactions within the chain and minimizes undesired interactions with the supporting substrate, has yet to be realized.

Here, our scanning tunneling microscopy/spectroscopy (STM/S) measurements and density functional theory (DFT) calculations jointly showed an exceptional category of 1D $CrCl_3$ atomic wires in a previously unobserved form. It was grown solely on an isotropic $NbSe_2$ vdW surface and was different from the nanoribbons of the 2D $CrCl_3$ phase that were also prepared in this work. Such a 1D wire, consisting of a single row of *face-sharing* $CrCl_6$ octahedra, is a large-gap semiconductor exhibiting a Néel-type antiferromagnetic (AFM) coupling. The narrowest 1D wire had a single unit-cell-width (denoted as a single-wire, SW) and could be stacked in parallel, forming vdW wire arrays. Interestingly, we observed that the STM tip triggered 1D-2D dimensional conversion, accompanied by a polymorphic phase transformation, in quad- or wider wires arrays. Our theory, along with experimental inputs, aided the understanding of the reason for the presence of the 1D phase and the subsequent 1D-2D conversion. The ability of the substrate to stabilize 1D-wire ends or 2D-flake edges by donating electrons through vdW couplings was of critical importance in controlling the dimensionality conversion and the phase transformation. Our findings refresh the understanding of vdW epitaxy and established an alternative idea for the vdW interface engineering of 1D nanostructures.

Figure 1a shows a typical STM image of $CrCl_3$ grown on a bulk $NbSe_2$ substrate with ~0.15 ML coverage (see the Methods Section for experimental details). The figure clearly indicates an anisotropic growth mode in which subnanometer-width or few-nanometer-width 1D structures developed along the armchair directions of the NbSe$_2$ lattice and were jointed at small 2D flakes. This 1D-2D coexisting feature appeared to be unique on the NbSe$_2$ substrate, in that an attempt of growth of CrCl$_3$ on a bilayer graphene (BLG) surface solely led to hexagonal-like flakes (Fig. 1e). These flakes showed the lattice of the well-known 2D CrCl$_3$ monolayer (Fig. 1f), namely, periodically spaced trimers of top-layer Cl atoms (marked by the dashed triangles) [12] with a separation of 6.0 Å, which denoted the 2D phase.

In terms of the 1D-2D complexes formed on the NbSe$_2$ substrate, the small flakes were also in the 2D phase (Supplementary Fig. S1), but the 1D wires seemed to be in a different phase. A close-up look at the wires (Fig. 1b) explicitly illustrates their composition of parallelly stacked integer numbers of single atomic wires. A further zoomed-in constant-current image of a single wire is shown in Fig. 1c. The figure exhibits a zigzag chain of protrusions with an apparent height of ~6.2 Å (Supplementary Fig. S2a, b), a periodicity ($c$) of 6.0 Å, and a separation of 1.8 Å (labelled as $d$ in Fig. 1c) between two rows of the zigzag chain. All these remarkable features of the observed 1D phase differentiated it from simply an ultranarrow ribbon of the 2D CrCl$_3$ monolayer, which was, again, confirmed by the higher topographic heights of the wires than those of the 2D flakes at certain sample bias (Supplementary Fig. S2c, d).

Thus, this unknown 1D CrCl$_3$ structure was referred to as the 1D phase. Its STM topographic features were analogous to those of the formerly predicted 1D polymorphic phase of CrBr$_3$ [26]. We thus proposed that the 1D phase structure (Fig. 2a) consisted of a single Cr$^{3+}$ row encapsulated by *face-sharing* CrCl$_6$ octahedra along the wire axis, showing a tube-like structure with six Cl lobes in the cross-sectional view (Fig. 2d). The 2D phase also contained the same individual octahedral units, but these units were connected to each other in an *edge-sharing* manner, as seen in the schematic model superposed in Fig. 1f. Therefore, the two phases have very different structural parameters, *e.g.*, the Cr-Cr spacing (2.95 Å *vs.* 3.44 Å), the Cr-Cr-Cr angle (180° *vs.* 120°), and the Cr-Cl-Cr bond angle (77° *vs.* 94°). Thus, discrepancies would likely exist in the exchange interactions and the magnetic ground states between these two structures.

The DFT calculated phonon spectra of the free-standing 1D $CrCl_3$ wire array (Fig. 2e) and the $CrCl_3$ single wire (Supplementary Fig. S3a) show no imaginary frequency, manifesting the structural stability. Upon the adsorption on the $NbSe_2$ surface, the wire prefers to sit on the surface through two Cl lobes at the site shown in Fig. 2b, leading to another two Cl atomic rows being directed towards the vacuum (Fig. 2d). These two Cl atoms made dominant contributions to the STM imaging, and the image therefore yielded a zigzag appearance like that shown in Fig. 1c. In addition, our calculations revealed the lattice constant $c$ = 5.96 Å, the Cl-Cl separation $d$ = 1.85 Å, and the apparent height of 6.20 Å, which were very consistent with the experimental values of 6.0 Å, 1.8 Å, and 6.2 Å. This good consistency was maintained in the wider wire arrays, *e.g.*, bi-wires (Fig. 1d and Fig. 2c, d). When two wires were stacked together, the experimental image (Fig. 1d) showed a different morphology from that of the single wire presented above, in that two rows of atoms in each zigzag chain appeared to have an apparent difference in height. This was consistent with the small tilt angle predicted by our calculations, as illustrated in the cross-sectional view (Fig. 2d, details in Supplementary Fig. S2e, f). In addition, for the interwire atomic registry, it was clearly seen in both the experimental and theoretical results that there was no shift along the wire axis between the two nearby zig-zag chains, and a rectangular unit-cell with an interwire separation of ~ 6.1 Å was thus formed.

Figure 2f shows typical differential conductivity (*dI/dV*) spectra taken for 1D SW/NbSe$_2$ and 2D flake/NbSe$_2$. The SW had an experimental bandgap of 2.32 eV, smaller than that of the 2D flake (2.68 eV). Our DFT calculations revealed that the SW prefers a Néel AFM state (Supplementary Figs. S3b and S4), showing a magnetic moment of ~2.8 $\mu_B$ primarily on each Cr atom and a bandgap of 1.25 eV (six-Cr finite wire) – 2.81 eV (infinite wire) depending on the wire length. Details of the band gap calculations are seen in Supplementary Table S1. The larger bandgap in the 2D monolayer was reproduced by our calculation in that the infinite form of the 2D phase gave a bandgap of 3.08 eV and the six-Cr 2D-phase flake gave a bandgap of 1.52 eV. The systematically smaller experimental values could be ascribed to the finite sizes of the 1D and 2D CrCl$_3$ structures measured in our experiments. The easy axis of the magnetic moments was perpendicular to the wire axis direction with a magnetic anisotropic energy of 0.02 meV/Cr, as extensively discussed in Supplementary Fig. S6.

The ABAB AFM order was examined by spin-polarized STM (SP-STM) measurements [27-28]. With out-of-plane magnetic fields, the bulk Ni tip was *in-situ* magnetized to be spin-up (spin-down) state before scanning [Methods]. The observations shown below were reproducible in our experiments using different Ni tips.

Figure 2h shows the constant current images of the same area taken with the spin-up (upper panel) and spin-down (middle panel) tip states. The bias here was chosen to be near the conduction band edge, which hosts the maximum polarization in the density of states (DOS) (Fig. S7). It is apparent in both images that a contrast appears between adjacent Cr sites. In addition, we observed that such contrast is reversed under flipping of the tip magnetization (up/down). For comparison, we show a spin-averaged image (lower panel in Fig. 2h) derived by adding the intensities of the two spin-resolved images, which resembles the topographic image taken with a Pt-Ir tip (Fig. 1c). The DFT simulated SP-STM images (Fig. 2i) are composed of dumbbell-shaped protrusions centred at alternate Cr atoms and oriented along the two nearest top-layer Cl atoms, which well reproduce the experimental results. Note that both the SP-STM imaging and the DFT calculations indicate the antiferromagnetic coupling in the SW. Our calculation results do not support the existence of spiral magnetism in the classical magnetism picture and at the DFT level (Fig. S4). However, the inherent quantum (thermal) fluctuation in quasi-1D systems and the small MAE of 0.02 meV between the *y* and *z* directions (arising from vdW adsorption) could result in complex magnetic long-range ordering, which appears to be another elusive topic that requires further exploration.

The formation of the 1D phase differentiated the NbSe$_2$ substrate from the others. Thus, it was interesting to more carefully examine the competition between the 1D and 2D phases on NbSe$_2$. Figure 3a–d illustrate a sequence of topographic images of CrCl$_3$/NbSe$_2$ as a function of the total coverage ($N_{total}$) ranging from 0.035 ML to 0.36 ML, which showed that the 1D phase was dominant at a lower coverage (0.035 ML,

Fig. 3a) and that the 2D phase became predominant for higher coverages (from 0.12 to 0.36 ML, Fig. 3b–d). This trend is more clearly depicted by a plot of the ratio ($R_{1D}$) between the partial coverage of the 1D wires $N_{wire}$ and $N_{total}$ as a function of $N_{total}$ (black dotted line in Fig. 3e) in which $R_{1D}$ continuously decrease. The plot of $N_{wire}$ (red dashed line) also supports the dominance of the 2D phase at higher coverages, because it first reaches a maximum at $N_{total}$ = 0.12 ML and then decreases with further deposition of $CrCl_3$. The reduced $N_{wire}$ is also accompanied by a narrowed average width of the 1D wires. Figure 3f plots the statistical width ($w$) distribution of the multi-wire arrays (MW) at two typical $N_{total}$ values, namely, 0.12 and 0.36 ML. Interestingly, an increase in $N_{total}$ reduced the most-probable width of the MW from $w$ = 4 to $w$ = 2, implying a likely spontaneous transformation of the 1D phase into the 2D phase when the MW width exceeded a threshold ($w_T$).

Furthermore, such 1D-2D dimensionality conversion and the threshold width $w_T$ were directly visualized in tip-controlled postgrowth transformations. Figure 3g shows that a moderate tip stimulus precisely induced a transformation from 1D wire-array to 2D nanoribbon (NR) at the nanometre scale (Methods Section). Through substantive experimental attempts, we found that the wider the array was, the more readily the transformation occurred, and the transformation never occurred for $w < 4$. For instance, in the right-most region of Fig. 3g, the MW with $w$ = 3 could retain their structure under all the considered tunneling conditions. Combining the as-grown and postgrowth transformations, the $w_T$ was estimated to be approximately three, which was well reproduced in our calculations as shown in Fig. S8. This threshold width represents the crossover of the formation energies of the two phases; hence, is crucial for comprehending the growth mechanism, as we will discuss in the following. Note that as the 2D phase was predicted to be in an XY FM state, the tip-induced structural transformations are, most likely, accompanied with changes of magnetism between 1D AFM and 2D FM states [14].

Our DFT results showed that the infinite 2D form of the $CrCl_3$ was 0.34 eV/Cr more stable than the infinite 1D form, which indicated that the edge-bulk energy competition of the 1D and 2D phases might play an essential role in stabilizing the 1D phase on the $NbSe_2$ substrate in the early growth stage. We thus focused on the relative stabilities of the SW and the 2D hexagonal flake (HF) containing the same finite number of Cr atoms, $n_{Cr}$ (see Supplementary Fig. S9 for their atomic models). We found that the bulk unit indeed has lower energy than the edge (end) unit in both phases (Supplementary Table S2). Note that the ratio between the numbers of end (edge) and bulk units in the SW is generally smaller than that in the HF, and the discrepancies in the ratio are $n_{Cr}$ dependent. For instance, with the smallest $n_{Cr}= 6$ (Fig. 4a, b), SW-6 contains two end units while all six Cr atoms in HF-6 are edge units, yielding the largest difference in the edge(end)/bulk ratio. This might have accounted for the superior stability of the SW in the early growth stage.

However, the inferred superior stability of SWs could not be illustrated by directly comparing the total energies of HF-6 and SW-6 because of the unequal numbers of Cl atoms. The introduction of the formation enthalpies, *i.e.*, the consideration of the difference in the Cl number and the Cl chemical potential $\mu_{Cl}$, solved this issue (see the Methods Section for details of the formation enthalpy $H$ and the total energies). Figure 4c plots the difference in the formation enthalpies of the HFs and their corresponding SWs for the Cl-rich ($\mu_{Cl}$ = 0.08 eV) and Cl-deficient ($\mu_{Cl}$ = −2.27 eV) limits (Methods Section). The SW was found to be more stable (with a positive $\Delta H$ value) for both limits when $n_{Cr}$ = 6. With increasing $n_{Cr}$, $\Delta H$ showed an overall declining tendency towards the negative regime. Although this declining trend implied a growth mode transformation from 1D wires to 2D flakes, these results could not explain the fact that the SW was solely formed on NbSe$_2$.

With the adsorption on both the NbSe$_2$ and BLG substrates, substantial charge transfer was identified in our calculation, especially at the ends (CrCl$_{4.5}$ units, Fig. 4d) or edges (CrCl$_4$ units, Fig. 4e). A total amount of 1.86 $e$ was transferred from NbSe$_2$ to SW-6, primarily at the six end Cl atoms (~0.3 $e$/Cl, Fig. 4d), which stabilized the end unit with a larger adsorption energy (Supplementary Table S2-S3). The delocalized π-orbital nature of the graphene electrons, along with the graphene work function and the energy levels of the SW (or HF) (Supplementary Fig. S10), led to an additional donated charge of 0.48 $e$ for SW-6, further lowering its energy of the end unit (Supplementary Table S2). In comparison with the SWs, the HFs had a stronger ability to accept electrons, and 0.42 $e$ and 0.50 $e$ more electrons were transferred from the NbSe$_2$ and graphene substrates to HF-6, respectively, mainly around the outermost Cl edge atoms. These additionally transferred electrons provided extra stabilization for the 2D edges of the HF on the graphene substrate (Fig. 4e), on which $E_{edge}^{HF}$ = −10.42 eV/Cr was surprisingly lower than $E_{bulk}^{SW}$ = −10.13 eV/Cr (Supplementary Table S2). This partially, or even fully broke the superior stability of the freestanding 1D phase in the early growth stage. Note that on the NbSe$_2$ substrate, the $E_{\text{edge}}^{\text{HF}} = -10.13$ eV/Cr remained higher than the $E_{\text{bulk}}^{\text{SW}} = -10.23$ eV/Cr, which favoured the possibility of 1D phase growth. See Supplementary Table S2 for comprehensive results. The stronger charge transfer at the end (edge) results from their electron acceptor nature (valence of 3+ for Cr), while the structural relaxation plays a minor and passive role (Fig. S11).

This inferred favoured growth was indeed reflected in the plot of $\Delta H$ between the HF and SW for $n_{\text{Cr}} = 6$ and 12 upon adsorption on the two substrates (Fig. 4f) for an experimentally estimated $\mu_{\text{Cl}}$. In the following, we first discuss this estimation of $\mu_{\text{Cl}}$. In practical experiments, the phase transformation mostly occurred from 1D MW to 2D NR, rather than directly to hexagonal flakes. We compared the $\Delta H$s between the MW and NR that formed on NbSe$_2$ with the equivalent width $w$ (see Fig. 4g and h for their atomic models with $w$ = 3 Cr rows). The phase diagram as functions of $w$ and $\mu_{\text{Cl}}$ is displayed in Fig. 4i, where a black dotted line represents the contour line of $\Delta H = 0$. According to the experimental fact that the most-probable width varied between $w_T = 2$ and 4 (Fig. 3f and 3g), we can estimate a range of experimental $\mu_{\text{Cl}} = -0.803$ to $-0.217$ eV/Cl (shadow region in Fig. 4i).

In Fig. 4f, the colour-filled bars represent the ranges of $\Delta H$ for the experimental range of $\mu_{\text{Cl}}$. It is indeed shown that the 1D phase retained more stability throughout the whole experimental range of $\mu_{\text{Cl}}$ on the NbSe$_2$, while on graphene, the possibility of favouring the 2D growth appeared. These results demonstrated that the selective choice of a substrate with a proper ability to donate electrons could tune the 1D and 2D growth modes of CrCl$_3$ on the substrate. This new mechanism is significantly distinct from the common notion of vdW epitaxy, in which the structure of the epilayer is nearly unrelated to the substrate [15].

In conclusion, we demonstrated the first experimental realization of single-unit-cell-width CrCl$_3$ atomic wires and their stacked few-wire arrays on the surface of NbSe$_2$. The single CrCl$_3$ wire was identified as a new 1D polymorphic phase of the 2D magnetic material CrCl$_3$ with a large bandgap and an antiferromagnetic ground state. Rich emergent quantum phenomena are anticipated in this unprecedented hybrid system consisting of vdW integrated 1D spin chain/2D superconductor. Moreover, a thorough understanding of the underlying growth mechanism revealed the elegant role of the electronic interactions at the vdW interface in controlling the dimensionality conversion and polymorphic phase transformation in epitaxial growth. This study expands the application scope of vdW interface engineering, offering an easy and flexible means to fabricate exotic 1D nanostructures.

**Methods Section**

**Growth of CrCl$_3$**: The CrCl$_3$ were grown on a freshly cleaved NbSe$_2$ substrate with compound source molecular beam epitaxy (MBE). Anhydrous CrCl$_3$ powder of 99% purity was evaporated from a Knudsen cell. The growth speed of ~0.02 ML/min was determined by checking the coverages of the as-grown samples. The NbSe$_2$ substrate was kept at room temperature during growth. The BLG/SiC substrate was synthesized by silicon sublimation from the (0001) plane (Si face) of n-type 6H-SiC [29]. The optimal substrate temperature for the growth of the CrCl$_3$ monolayer flakes was ~500

K. Below this temperature, the CrCl$_3$ tended to form 2D fractal structures features on the bilayer graphene surface (for details, see Supplementary Fig. S12).

**STM/S Measurements**: After sample preparation, the sample was inserted into a low-temperature scanning tunneling microscope (STM, Unisoku USM-1300) housed in the same ultra-high vacuum system. Polycrystalline Pt-Ir STM tip was used in our experiments. The bias voltage was applied to the sample. All STM images presented in this article were taken at 4.3 K. The d$I$/d$V$ spectra were measured by using the lock-in technique with reference signal at 963 Hz. The modulation amplitudes were set as 50 μV (Fig. 2e) and 10 mV for all other spectra. The SP-STM measurements were carried out via electrochemically etched Ni tips in a constant-current mode [30, 31]. The spin-polarization of the Ni tip was calibrated on Co/Cu(111) (Supplementary Fig. S13). Following the procedure reported in ref. [32-34], perpendicular magnetic fields of + 0.7 T (− 0.7 T) were used to magnetize the ferromagnetic Ni tip acquiring the spin-up (spin-down) tip state. The external fields were usually held for five minutes and then slowly reduced to zero. The magnetized tip was then used to perform STM imaging at 0 T on the same location.

**Tip-induced Phase Transformation**: The structural phase of the stacked wire arrays was not affected by imaging using a sample bias 0 V > $V_B$ > −2.5 V. Once the sample bias was in excess of ∼−2.8 V, the transformation to the 2D phase could easily occur. The practical procedure we adopted for Fig. 3g was as follows: running continuous scanning of the selected region at the constant height mode with the initial scanning parameters of $V_B$ = −2.9 V, $I$ = 3 pA until an abrupt drop of $I$ (usually down to zero) was observed (since the 2D phase was slightly lower in terms of the topographic height).

**DFT Calculations**: Calculations were performed using the generalized gradient approximation in the Perdew-Burke-Ernzerhof (PBE) form [35] for the exchange-correlation potential, the projector augmented wave method [36], and a plane-wave basis set as implemented in the Vienna ab-initio simulation package (VASP) [37]. Dispersion corrections were made at the van der Waals density functional (vdW-DF) level [38] with the optB86b functional for the exchange potential (optB86b-vdW) [39] in all structural relaxations. The structures were fully relaxed until the residual force per atom was less than 0.005 (0.02) eV/Å for free-standing (substrate-supported) SWs and HFs. On-site Coulomb interactions on the Cr $d$ orbitals were considered using a DFT+U method [40] with $U$= 3.9 eV and $J$ =1.1 eV, consistent with the values used in the literature [41]. A vacuum layer larger than 15 Å was used in all supercells to avoid interactions between the slabs of adjacent supercells. An energy cut-off of 700 (400) eV was used for the plane wave basis set in calculating free-standing (substrate-supported) structures. A $k$-mesh of 14×1×1 was used to sample the first Brillouin zone of freestanding infinite SWs while the Gamma point was used in other calculations.

The PBE functional was used in comparison of relative energies of individual configurations based on the atomic structures optimized using the optB86b-vdW functional. Spin-orbit coupling was considered in all energy comparisons. The accuracy of such set of methods were well tested in calculations on many 2D magnets [42-49] The charge transfer between the SWs (HFs) and the substrate was evaluated using the Bader charge analysis method [50]. The charge variation of a Cl atom was defined as the Bader charge of the Cl atom in SW-6 (HF-6) being placed on the substrate minus the that in the free-standing form. The phonon dispersion was calculated using the density functional perturbation theory, as implemented in the PHONOPY code [51].

The periodic direction of the multi-wires array (zig-zag direction of corresponding nanoribbon) is oriented along the arm-chair direction of NbSe$_2$, which was consistent with the experimental observation. For the finite SWs and HFs with a NbSe$_2$ or graphene substrate, we used the same relative orientations as the periodic case. Two-layer NbSe$_2$ (graphene) was used to model the substrate, in which the bottom layer was kept fixed and the top layer was allowed to fully relax.

**Calculation of formation enthalpy *H***: The formation enthalpy $H$ for finite SW and HF is defined as $H = (E_{total}^{HF/SW} - n_{Cl} \cdot \mu_{Cl} - n_{Cl} \cdot \mu_{Cl})/n_{Cr}$. $E_{total}^{HF/SW}$ is the total energy of SW or HF, $n_{Cr}(n_{Cl})$ is the number of Cr (Cl) atoms. Chemical potentials of Cr and Cl in CrCl$_3$ fulfill the equation $\mu_{Cr} + 3\mu_{Cl} = \mu_{Cr}^* + 3\mu_{Cl}^* + \Delta H_{CrCl_3}$, where $\mu_{Cr}^*$ is the chemical potential of Cr in the bulk form, $\mu_{Cl}^*$ is the chemical potential of Cl$_2$ and $\Delta H_{CrCl_3}$ is the formation energy of CrCl$_3$. At the Cl rich limit, one gets $\mu_{Cl} = \mu_{Cl}^*$, while at Cl deficient limit, $\mu_{Cl} = \mu_{Cl}^* + \Delta H_{CrCl_3}/3$.

The total energy reads $E_{total}^{HF/SW} = n_{edge}^{HF/SW} \times (E_{edge-fs}^{HF/SW} + E_{edge-ad}^{HF/SW}) + n_{bulk}^{HF/SW} \times (E_{bulk-fs}^{HF/SW} + E_{bulk-ad}^{HF/SW}) + E_{SEC}^{HF/SW}$. If we set $E_{edge}^{HF/SW} = E_{edge-fs}^{HF/SW} + E_{edge-ad}^{HF/SW}$ and $E_{bulk}^{HF/SW} = E_{bulk-fs}^{HF/SW} + E_{bulk-ad}^{HF/SW}$, it further reads $E_{total}^{HF/SW} = n_{edge}^{HF/SW} \times E_{edge}^{HF/SW} + n_{bulk}^{HF/SW} \times E_{bulk}^{HF/SW} + E_{SEC}^{HF/SW}$, where subscripts "edge" and "bulk" refer to edge and bulk units of SW and HF, postfixes "-fs" and "-ad" denote free-standing and adsorbed forms of the CrCl$_x$ units, $n$ is the number of CrCl$_x$ units in a finite size SW or HF, subscript "SEC" stands for spin-exchange coupling.

In order to determine the relative stability of the SW and HF phases with the same $n_{Cr}$, we could examine the difference of normalized enthalpy $\Delta H_{\text{HF-SW}}=[(n_{\text{edge}}^{\text{HF}} \times E_{\text{edge}}^{\text{HF}} + n_{\text{bulk}}^{\text{HF}} \times E_{\text{bulk}}^{\text{HF}}) - (n_{\text{edge}}^{\text{SW}} \times E_{\text{edge}}^{\text{SW}} + n_{\text{bulk}}^{\text{SW}} \times E_{\text{bulk}}^{\text{SW}}) + (E_{\text{SEC}}^{\text{HF}} - E_{\text{SEC}}^{\text{SW}}) + \Delta n_{\text{Cl}} \cdot \mu_{\text{Cl}}]/n_{\text{Cr}}$, where $\Delta n_{\text{Cl}}$ indicates the number difference of Cl atoms between the SW and HF phases. A $\Delta H_{\text{HF-SW}}$ value larger (smaller) than zero indicates the superior stability of SW (HF). Here, the magnetic term is considerably smaller than other terms and is thus negligible in comparing relative energy of SW and HF. Number $n_{\text{edge}}^{\text{SW}}$ always equals to 2 in a finite SW and number $n_{\text{bulk}}^{\text{HF}}$ is often smaller than $n_{\text{edge}}^{\text{HF}}$ or even approaches to zero in a small-size HF. We could consider a simplified and qualitative relation $\Delta H_{\text{HF-SW}} \sim E_{\text{edge}}^{\text{HF}} - E_{\text{bulk}}^{\text{SW}} + (\Delta n_{\text{Cl}} \cdot \mu_{\text{Cl}})/n_{\text{Cr}}$ to judge the relative stability of SW and HF by assuming $E_{\text{edge}}^{\text{HF}} \approx E_{\text{bulk}}^{\text{HF}}$ and $E_{\text{bulk}}^{\text{SW}} \approx E_{\text{edge}}^{\text{SW}}$. Thus, if $E_{\text{bulk}}^{\text{SW}}$ is lower than $E_{\text{edge}}^{\text{HF}}$, we could have chance to obtain a positive $\Delta H_{\text{HF-SW}}$ and thus more stable SW in a certain $\mu_{\text{Cl}}$ range as derived from the experiments.

**Data availability**

The data that support the findings of this study are available within the article and its Supplementary Information. The source data are available from the corresponding authors upon request.

**Code availability**

The code that supports the findings of this study is available from the corresponding author upon reasonable request.

**Acknowledgments**

We thank the supports from the National Key R&D Program of China (Grant No. 2018FYA0305800, 2018YFA0703700 and No. 2018YFE0202700), the National Natural Science Foundation of China (Grant No. 11974012, 12134011, 61888102, 21802165, 11974422 and 12104504), the Strategic Priority Research Program of Chinese Academy of Sciences (Grant No. XDB30000000), the Fundamental Research Funds for the Central Universities, China, and the Research Funds of Renmin University of China [22XNKJ30 (W.J.)]. D.P.G. was supported by the Outstanding Innovative Talents Cultivation Funded Programs 2022 of Renmin University of China. S.Z.L. and C.W were supported by the China Postdoctoral Science Foundation (2019M652694 and 2021M693479). Calculations were performed at the Physics Lab of High-Performance Computing of Renmin University of China, Shanghai Supercomputer Center.

**Author Contributions**

S.Z.L., Z.B.C. and Y.P.G. carried out the MBE growth and STM/S measurements. D.P.G. and J.W. performed the first-principles calculations. J.H.D. and Y.S.B. contributed to the sample preparation and STM measurements. Z.B.C. completed the SP-STM characterization. C.T. and Y.G.S coordinated the preparations of $NbSe_2$. C.W. and L.W.Z participated in discussions of theoretical calculations. J.H. and C.D.Z. initiated and coordinated the work, participated in the experiments, analyzed data. W.J. conceived the theoretical calculations and analysis. S.Z.L., D.P.G., J.H., J.W. and C.D.Z. wrote the manuscript with inputs from the other co-authors.

**Competing interests**

The authors declare no competing financial interests.

**Additional information**

**Supplementary information** The online version contains supplementary material available at https://doi.org/

**Figure Captions**

**Figure 1 | STM images of 1D CrCl$_3$ wires on NbSe$_2$ and 2D CrCl$_3$ flakes on BLG. a**, STM image of CrCl$_3$ grown on a bulk NbSe$_2$ surface (1.1 V, 10 pA). The inset shows the atomically resolved STM image taken on the nearby NbSe$_2$. The prime vector $a_{sub}$ of the NbSe$_2$ lattice is marked as shown. **b**, Close-up image (1.1 V, 10 pA) showing that the as-grown 1D structures consisted of integer numbers of atomic wires, which were jointed at the small flakes. **c, d**, Atomically resolved images (0.5 V, 100 pA) of the isolated CrCl$_3$ single wire and bi-wires. The schematics of the atomic models are superposed on the images. Only the topmost Cl atoms (yellow in the model) were imaged, with the formation of the zigzag chains marked by the red dashed lines. The rectangular unit cell of the bi-wires is marked by black dashed lines. **e**, Typical image (1.6 V, 5 pA) of the sample grown on BLG showing only 2D flakes with a monolayer thickness. **f**, Atomically resolved image of the 2D CrCl$_3$ phase in **e** with an overlaid schematic model. The dashed black triangles indicate the Cl trimers. The small 2D flakes in **a** show the same 2D phase (Supplementary Fig. S1).

**Figure 2 | Atomic, electronic, and magnetic structures of 1D CrCl$_3$ wires. a**, Perspective view of a single CrCl$_3$ wire composed of face-sharing CrCl$_6$ octahedra (shadowed). **b, c**, Top views of the most stable configurations of the single wire and bi-wires on the adsorbed NbSe$_2$ obtained in our DFT calculations. The corresponding cross-sectional views are shown in **d**. The Cr atoms are displayed in blue, and the Cl atoms are displayed in yellow (for the top-most ones) and green. **e**, Calculated phonon spectrum of the CrCl$_3$ wire array with an infinite width. **f**, Typical d$I$/d$V$ spectra (on a logarithmic scale) of a single wire (orange) and a 2D CrCl$_3$ flake (blue). The bandgap values are labelled. **g**, Schematics of the spin configuration in a single wire. A total energy comparison among various magnetic configurations indicates that the most stable configuration is the Néel AFM (ABAB) state (Supplementary Fig. S4). The arrows represent the spin directions. The lower panel shows the visual spin density distribution. (red: spin-up; blue: spin-down) **h**, Constant current images taken with a spin-polarized Ni tip. The upper and middle panels show the same area imaged by the tip magnetized with +0.7 T and -0.7 T magnetic fields, respectively. Both images were acquired at +1.5 V and 5 pA. Cross-check of the contrast reversal is seen in Supplementary Fig. S5. The lower panel is a sum of the above two panels. Scale bars are 0.5 nm. The vertical grid lines represent the Cr sites (blue dots), and atomic models are superimposed on each panel. **i**, Simulations of STM images with pure spin-up/down sample DOS at +1.0 V.

**Figure 3| Coverage-dependent and tip-manipulated 1D-to-2D transformations. a–d**, Coverage-dependent topographic images of CrCl$_3$ grown on NbSe$_2$ with the total coverage ranging from 0.035 ML to 0.36 ML. Scale bars are 50 nm. **e**, Plot of the coverage $N_{wire}$ of the 1D wires and the coverage ratio $R_{1D}$ = $N_{wire}/N_{total}$ as functions of $N_{total}$. **f**, Statistical analysis of the widths of the wire arrays at $N_{total}$ = 0.12 ML and 0.36 ML. **g**, Tip-manipulated transformations from the 1D multi-wire arrays (MWs) to 2D

nanoribbons (NRs). The technical details are discussed in the Methods Section. Images were taken at −0.6 V and 10 pA; scale bars are 2 nm.

**Figure 4 | Theoretical understanding of the 1D phase growth and the 1D-2D transformation. a**, Side view of SW-6. **b**, Top view of HF-6. The upper and lower surface Cl atoms are represented by yellow and green balls, respectively. **c**, Plot of $\Delta H$ between the free-standing SW and HF at the Cl-rich (green square) and Cl-deficient (green circle) extremes as a function of $n_{Cr}$. The HFs with zigzag (ZZ) edges ($n_{Cr}$ = 6, 24 and 54) are more stable than the arm-chair (AC) ones ($n_{Cr}$ =12 and 36). **d**, **e**, Plots of the charge variations for each individual Cl atom in SW-6 (**d**) and HF-6 (**e**) on the NbSe$_2$ (orange) and graphene (blue) substrates. The indices of the Cl atoms are labelled in **a** and **b**. The edge atoms are marked by grey shadows. **f**, $\Delta H$ between SWs and HFs (at $n_{Cr}$ = 6 and 12) with the substrates involved. The colour-filled bars represent the estimated range of experimental $\mu_{Cl}$. The circle and square symbols represent the results at the Cl-deficient and Cl-rich limits, respectively. **g**, **h**, Atomic structures of multi-wire array (MW, $w$ = 3) on NbSe$_2$ and its corresponding nanoribbon (NR). **i**, Phase diagram of $\Delta H$ as functions of $w$ and $\mu_{Cl}$. The contour line of $\Delta H = 0$ is plotted as the dashed line. The shadow region corresponds to the experimental observation of $w_T$ = 2–4.

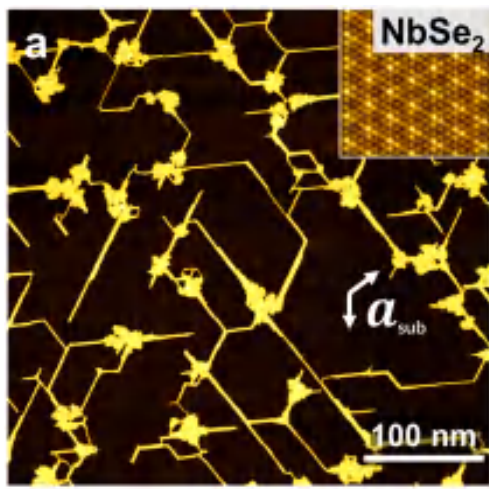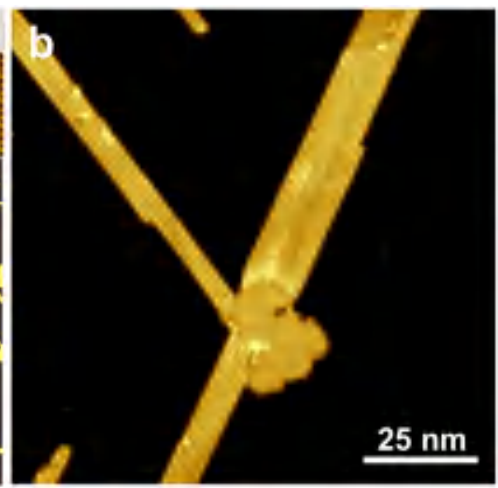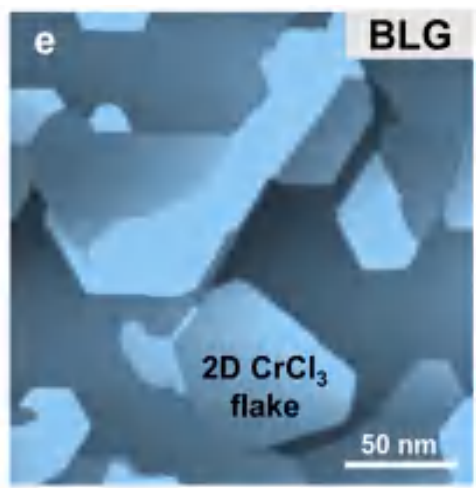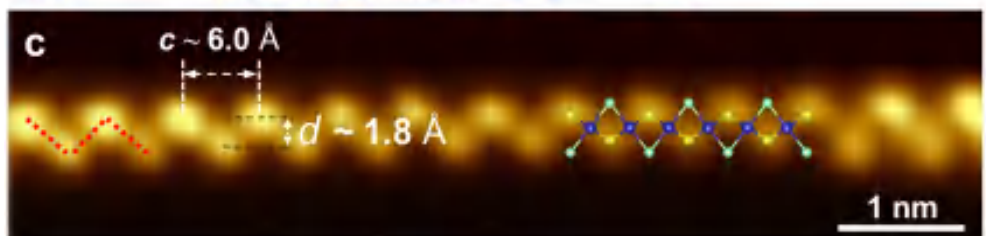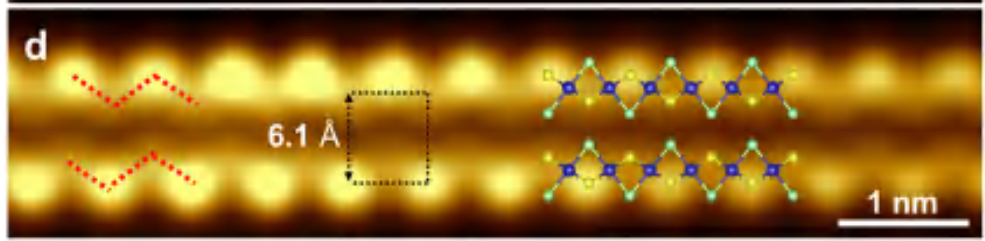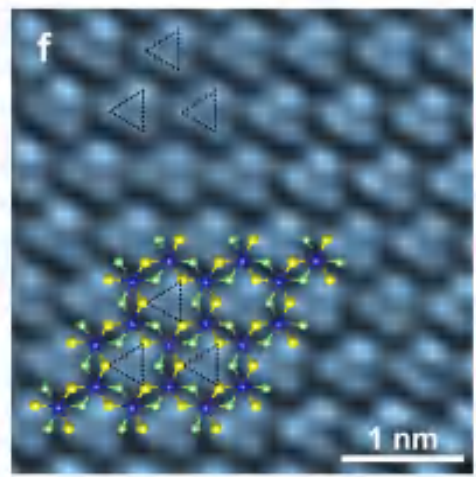

1D CrCl$_3$     2D CrCl$_3$

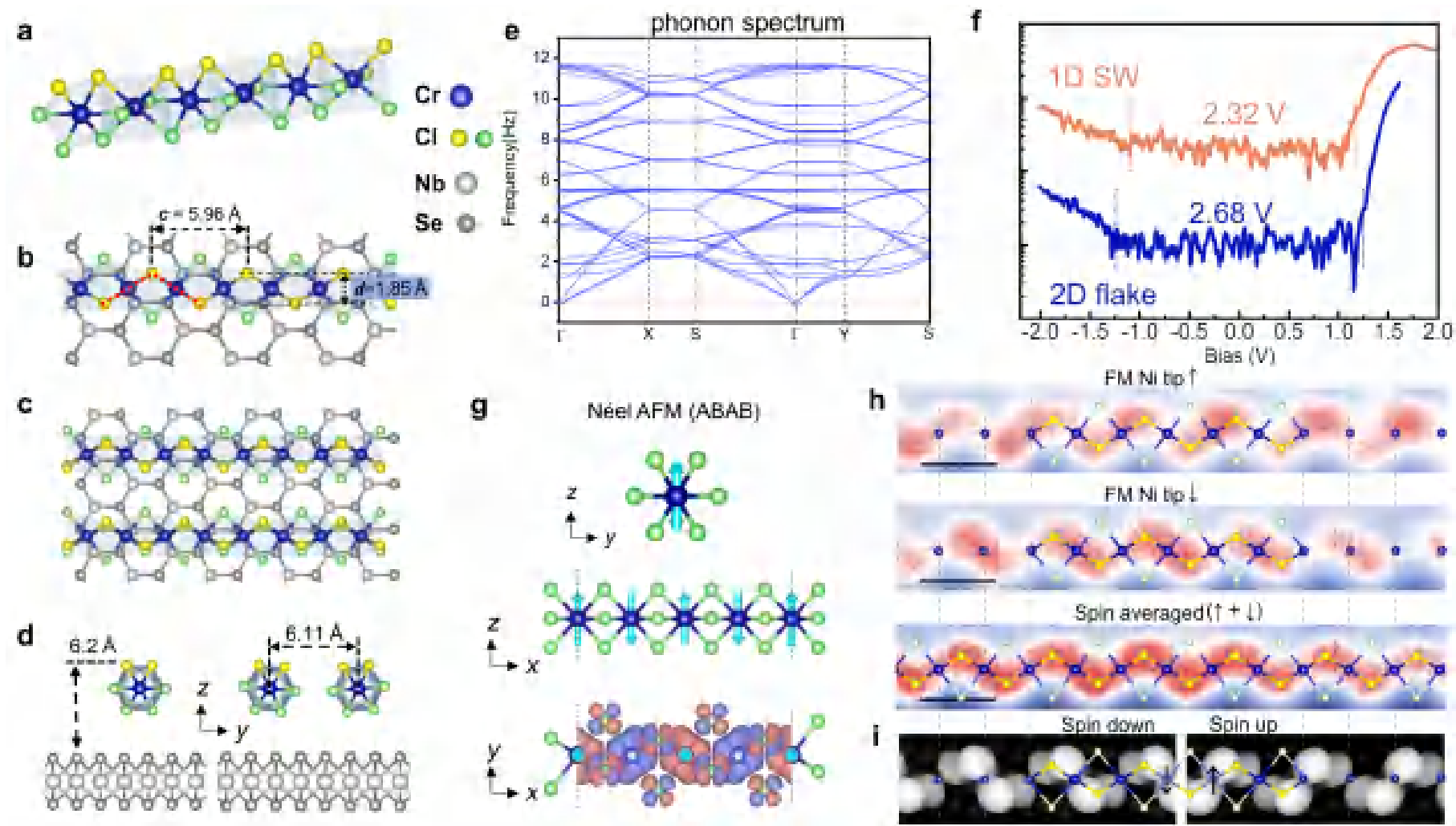

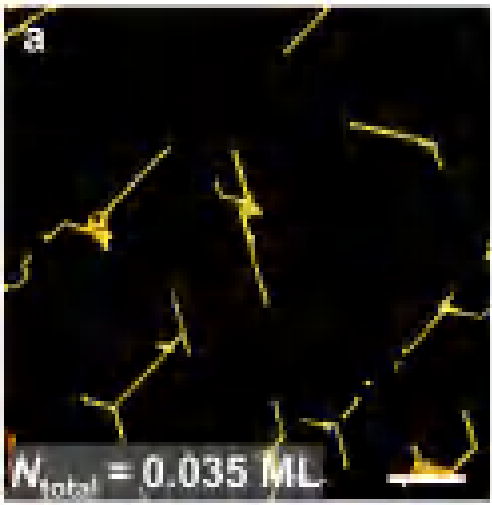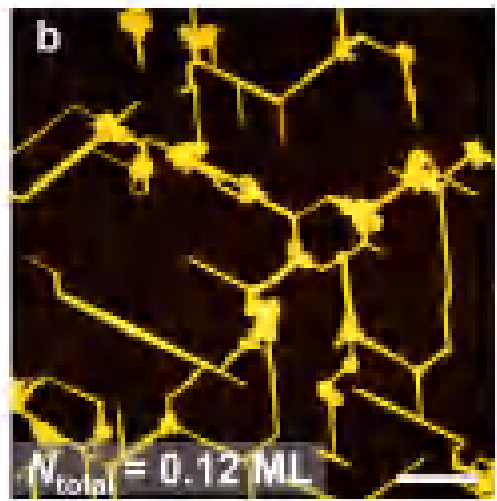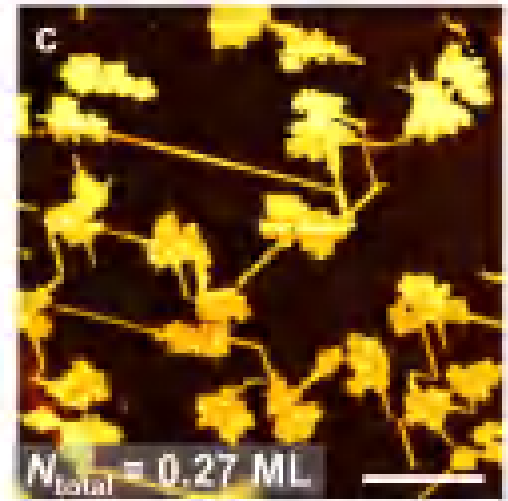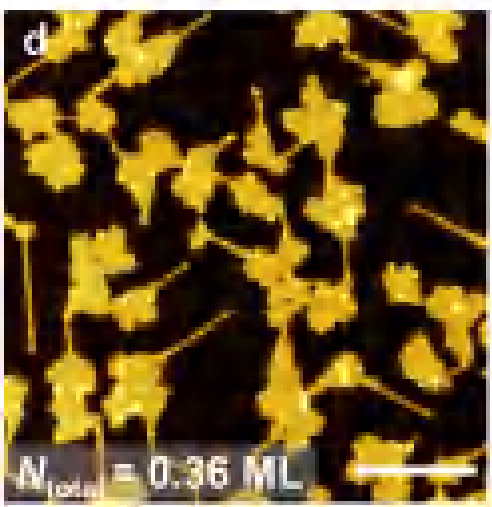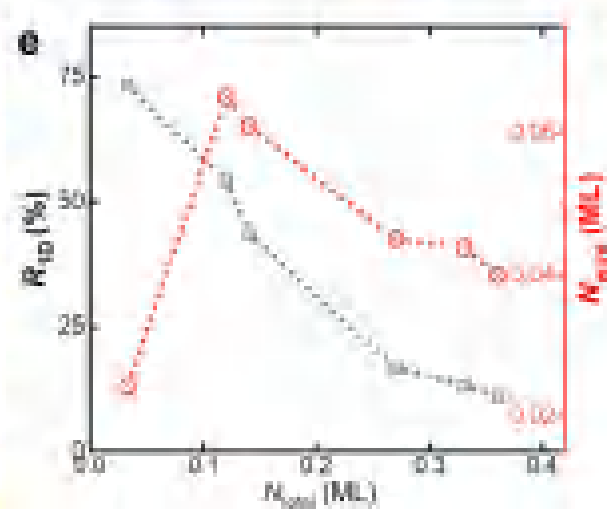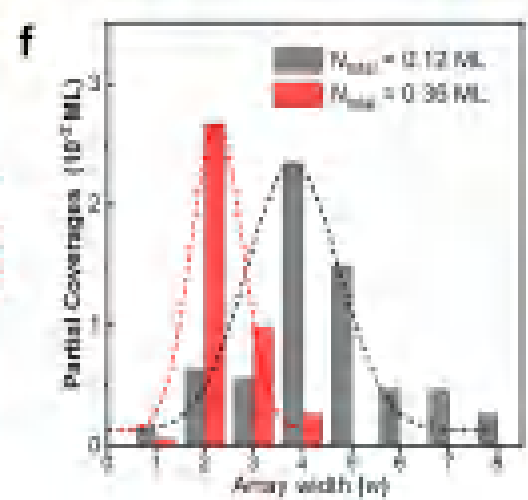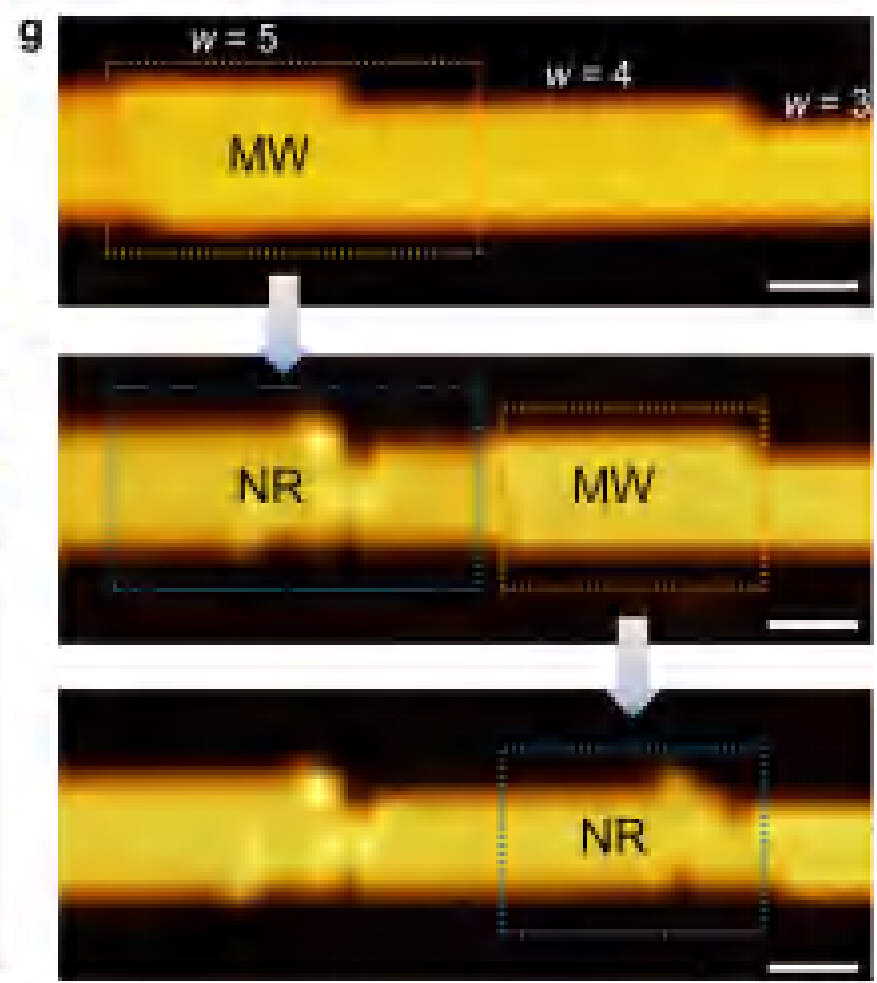

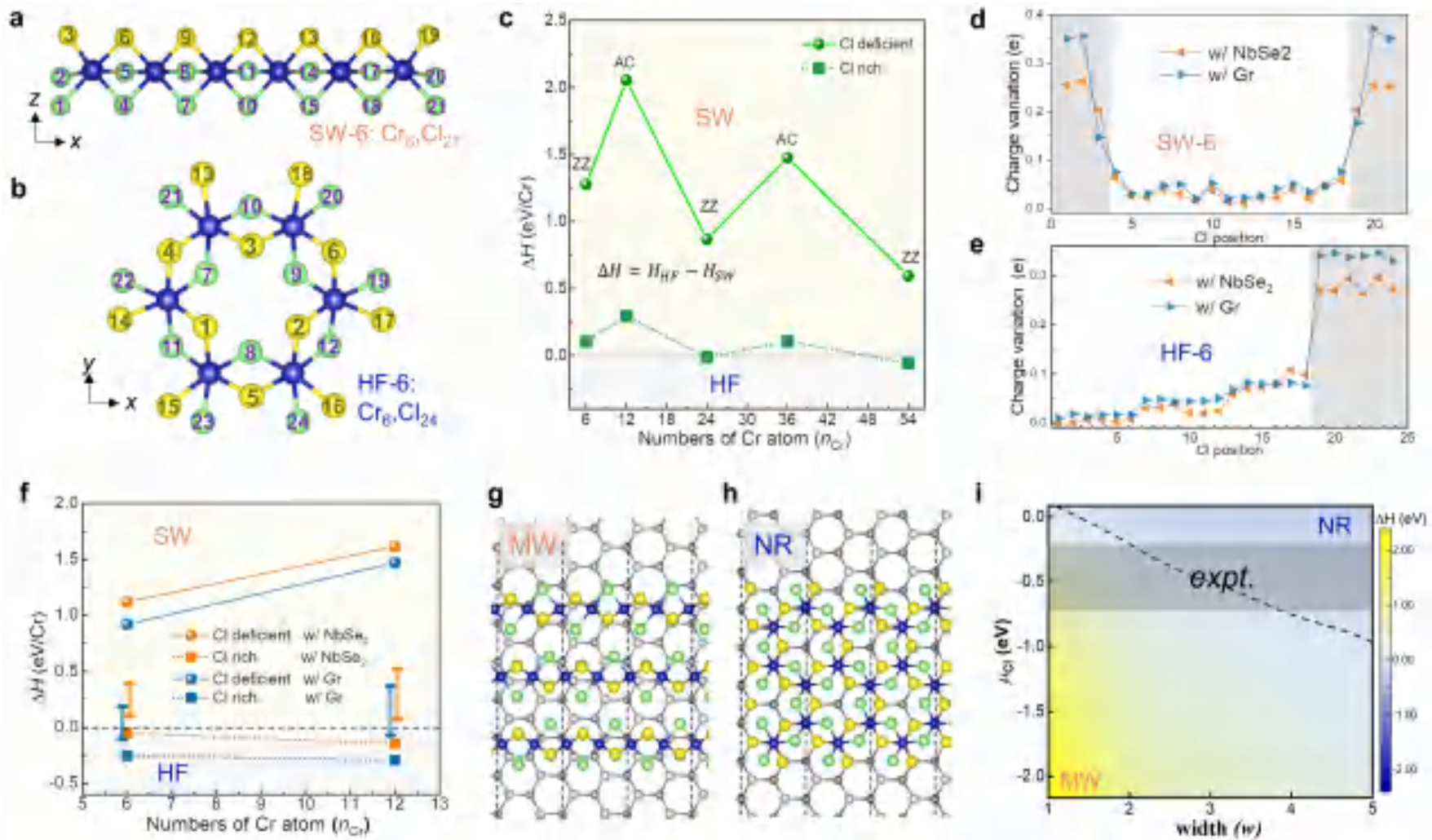